\documentclass[prl,twocolumn,superscriptaddress]{revtex4}
\usepackage{graphicx}
\usepackage{amsmath,graphicx}
\usepackage{epsfig}
\usepackage{paralist}
\usepackage{amsmath,amssymb,mathrsfs}
\usepackage{hyperref}
\usepackage{paralist}
\usepackage{underscore}
\usepackage{natbib}

\begin{document}

\title{Chiral d-wave superfluid in periodically driven lattices}

\author{Shao-Liang Zhang, Li-Jun Lang, Qi Zhou}
\affiliation{Department of Physics, The Chinese University of Hong Kong, Shatin, New Territories, HK}

\date{\today}

\begin{abstract}
{\bf Chiral d-wave superfluid is a preliminary example of topological matters that intrinsically encodes interaction effects. It exhibits fascinating properties including a finite Chern number in the bulk and topologically protected edge states, which have been invoking physicists for decades. However, unlike s-wave superfluids prevalent in nature, its existence requires a strong interaction in the d-wave channel, a criterion that is difficult to access in ordinary systems. So far, such an unconventional superfluid has not been discovered in experiments. Here, we present a new principle for creating a two-dimensional chiral d-wave superfluid using periodically driven  lattices. Due to an imprinted two-dimensional pseudospin-orbit coupling, where the sublattice index serves as the pseudospin, s-wave interaction between two hyperfine spin states naturally creates  a chiral d-wave superfluid. This scheme also allows physicists to study the phase transition between the topologically distinct s- and d-wave superfluids by controlling the driving field or the particle density.}
\end{abstract}

\maketitle

Superfluidity takes center stage in modern condensed matter physics\cite{BCS, Leggett}. Whereas a large number of elements on the periodic table become superfluids or superconductors at low temperatures, 
most of them are conventional $s$-wave ones. Though superfluidity could also occur in high-partial-wave channels, such as 
 $p$-wave superfluids in Helium and $d$-wave superconductors in cuprates \cite{psf3he,dsf}, it is very rare that a high-partial-wave superfluidity is accompanied by the time-reversal symmetry breaking. Such chiral superfluids, for instance, the $p_x+ip_y$ and $d_{x^2-y^2}+id_{xy}$ superfluids \cite{pandip,randg,Yao, Volovik,Laughlin,  Senthil}, give rise to exotic phenomena in an interacting many-body system. It has been realized that one need go beyond the conventional symmetry breaking paradigm to describe these chiral superfluids, since they are distinguished from their counterparts that respect the time-reversal symmetry, such as $p_x$ and $d_{x^2-y^2}$, by a finite Chern number, a topological property in the bulk, and chiral edge states protected by topology in a finite system. There have been experimental evidences for the $p_x+ip_y$ superconductor, which exhibits a Chern number $C=1$, to exist in $Sr_2RuO_4$ \cite{MandM}. For the chiral $d_{x^2-y^2}+id_{xy}$ superfluid with a Chern number $C=2$, there has been no experimental observation yet.

To realize a chiral $d_{x^2-y^2}+id_{xy}$ superfluid, theoretical studies have proposed using doped graphene for enhancing the $d$-wave interaction \cite{dopedC}, or magnetic impurities for mixing $d_{xy}$ with $d_{x^2-y^2}$ in cuprates \cite{Laughlin,magimp}. These proposals requiring sophisticated doping of solid materials have not been realized in experiments so far. In ultracold atoms, though the interaction can be tuned by Feshbach resonance,  the strong atom loss near  high-partial-wave resonance makes it challenging to explore chiral superfluids with a large $d$-wave scattering length\cite{Jin, Cheng}. Here, we propose a new scheme that simply requires $s$-wave interaction. The idea is to engineer a special band structure in a lattice by breaking the time-reversal symmetry via a periodic driving field. Such a driving field imprints a two-dimensional pseudospin-orbit coupling in the single-particle band structure  $\sim k_y m_x+ k_xm_y$, where ${\bf k}=(k_x,k_y)$ is the momentum, and the sublattice index plays the role of a pseudospin-1/2 ${\bf m}$. Uploading two hyperfine spin states $\sigma=\uparrow, \downarrow$ that experience the same driving field to this lattice, the $s$-wave interaction between these hyperfine spin states inevitably induces a synthetic $d$-wave one, and the relative strengths of the interactions in the $s$- and $d$-wave channels can be well controlled by the driving field and other microscopic parameters. Since the time-reversal symmetry is readily broken in the single-particle level, when the synthetic $d$-wave interaction is dominant, a chiral $d$-wave superfluid naturally emerges. \\

{\bf Single-particle band structure}  We consider a two-dimensional lattice subject to a circularly polarized time-periodic gauge field ${\bf A}(\tau)=(A\sin \omega \tau, -A\cos \omega \tau)$. The Hamiltonian in the Floquet framework can be written as
\begin{equation}
H_0-i\hbar\partial_\tau=\frac{(\hat{\bf P}- {\bf A}(\tau))^2}{2M}+V({\bf r})-i\hbar\partial_\tau, \label{Hp}
\end{equation}
where ${\bf r}=(x,y)$, and $V({\bf r})$ is the lattice potential. This Hamiltonian applies to both electrons in solids and ultracold atoms with synthetic gauge fields \cite{drivenoptlat}.  By a simple transformation $x\rightarrow x+f\cos \omega \tau, y\rightarrow y+f\sin \omega \tau, \tau\rightarrow \tau $, one sees that equation (\ref{Hp}) is equivalent to
\begin{equation}
H_0-i\hbar\partial_\tau=\frac{\hat{\bf P}^2}{2M}+V(x+f\cos \omega \tau, y+f\sin \omega \tau)-i\hbar\partial_\tau, \label{Hs}
\end{equation}
where $f=A/(M\omega)$, except for a trivial position independent constant. Equation (\ref{Hs}) describes a shaken optical lattice that has attracted a lot of interest recently in ultracold atom physics \cite{shaken1d,haldane, Zhang, Zhou}. Our scheme therefore works also for shaken optical lattices.
\begin{figure}[tbp]
\begin{center}
\includegraphics[width=3.5in]{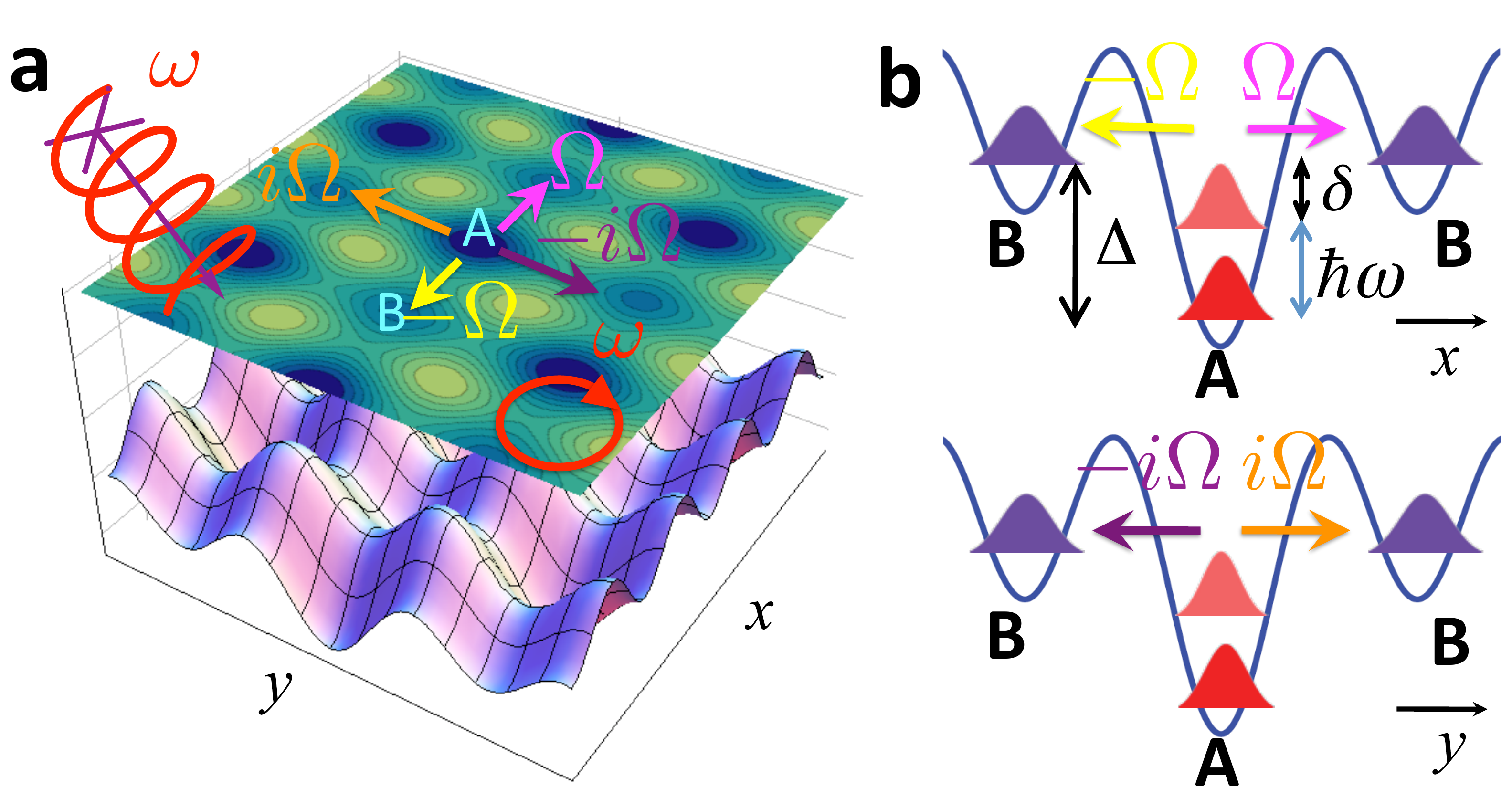}
\end{center}
\caption{\textbf{Periodically driven checkerboard lattice}. (\textbf{a}) Applying a circularly polarized light (red helix) or a circular shaking (red circle) to the checkerboard lattice, the tunneling from a A sublattice site to its four nearest B sublattice sites becomes dependent on the direction, i.e., $\sim \pm \Omega$ and $\sim \pm i \Omega$ along the $x$ and the $y$ directions respectively. $\Omega$ is the strength of the photon-assisted inter-sublattice tunneling.  (\textbf{b}) Red and blue clouds represent the Wannier wave functions in the A and B sublattice sites respectively. $\Delta$ is the energy offset of the two sublattices, and $\delta=\Delta-\hbar \omega$ is the one-photon detuning. }
 \end{figure}
Whereas our scheme is general, to concrete the discussion, we focus on a checkerboard lattice composed of A and B sublattices,
\begin{equation}
V({\bf r})=-V_0\left(\cos^2(\frac{\pi x}{d})+\cos^2(\frac{\pi y}{d})+2s \cos(\frac{\pi x}{d})\cos(\frac{\pi y}{d})\right), \label{cl}
\end{equation}
where $V_0$ is the lattice depth, $d$ is the lattice spacing, and $s$ is a tunable constant, as shown in figure 1(a). A strong lattice potential $V_z(z)$ is applied along the $z$ direction for creating a two-dimensional system. Whereas a checkerboard lattice is common in solids, such an optical lattice can also be created for ultracold atoms \cite{square,checkerboard}. The band structure of both the static  and the dynamically driven lattices can be solved exactly using the plane-wave expansion and the Bloch-Floquet theory respectively. For the static lattice, the lowest two bands are well separated from other ones for large $V_0$. The main contributions to these two bands come from A and B sublattices respectively, since a finite $s$ produces an energy offset  $\Delta$ between the nearest neighbor sites and suppresses the inter-sublattice tunneling, as shown in figure 1(b). However, once such a lattice is periodically driven, the atomic orbital in A(B) sublattice sites could absorb(emit) a photon and overcome the energy mismatch. Such a photon-assisted tunneling produced by the ${\bf A}(\tau)\cdot \hat{\bf P}$ term hybridizes the lowest two bands if $\omega$ is close to their band gap, and  the Hamiltonian in the rotating wave approximation is written as
\begin{equation}
K=\left(\begin{array}{cc}
\epsilon_{A{\bf k}} & \Omega_{\bf k}e^{i\varphi_{\bf k}} \\ \Omega_{\bf k}e^{-i\varphi_{\bf k}} & \epsilon_{B{\bf k}}
\end{array}\right),\label{K}
\end{equation}
where $\epsilon_{A{\bf k}}=4t_A\cos k_xd\cos k_yd+2t'_A(\cos 2k_xd+\cos 2k_yd)-\delta$, $\epsilon_{B{\bf k}}=4t_B\cos k_xd\cos k_yd+2t'_B(\cos 2k_xd+\cos 2k_yd)$, $\Omega_{\bf k}e^{i\varphi_{\bf k}}=2\Omega(i\sin k_xd-\sin k_yd)e^{-ik_xd}$ and $\varphi_{\bf k}=\arg\{(i\sin k_xd-\sin k_yd)e^{-ik_xd}\}$. $t_A$ ($t_B$) and $t'_A$ ($t'_B$) characterize the nearest- and next-nearest neighbors tunnelings in the A(B) sublattice,  $\Omega$ is the amplitude of the photon-assisted intra-sublattice tunneling, and $\delta=\Delta-\hbar\omega$ is the one-photon detuning. Though a finite $s$ doubles the unit cell of the square lattice with $s=0$, we have defined $k_x$ and $k_y$, as shown in figure 2,  to be consistent with the expression of the lattice potential shown in  equation (\ref{cl}).
\par
Equation (\ref{K}) has a clear physical interpretation in the real space. As shown in figure 1(a), the inter-sublattice tunneling carries a nontrivial phase factor in the tight-binding model.  For the tunneling from a $A$ sublattice site to its four nearest neighboring B sublattice sites, the tunneling is proportional to $1, i, -1, -i$ respectively. These complex tunnelings come from the circularly driving field that break the time-reversal symmetry. The relevant basis wave functions are $W_A({\bf r})e^{i\omega \tau}$ and $W_B({\bf r})$, in the language of Floquet theory, where $W_A({\bf r})$ and $W_B({\bf r}) $ are the Wannier functions of the A and B sublattice respectively. Physically, the phase factor $e^{i\omega \tau}$ corresponds to absorbing one photon in the A sublattice. The tunneling from A to B sublattice sites along the $\hat{x}$ and $-\hat{x}$ directions, which are referred to as  $t_{\pm \hat{x}}$, are proportional to $\int_0^{\frac{2\pi}{\omega}}d\tau\int d{\bf r} W_B({\bf r}\pm d\hat{x})  \sin(\omega \tau) \hat{P}_x (W_A({\bf r})e^{i\omega \tau})$, where $\hat{x}$ is the unit vector. Since both $W_A({\bf r})$ and $W_B({\bf r})$ have even parity, one naturally sees that  $t_{\pm \hat{x}}\sim \pm \Omega$. As for the inter-sublattice tunneling along the y direction,  $t_{\pm \hat{y}}\sim -\int_0^{\frac{2\pi}{\omega}}d\tau\int d{\bf r} W_B({\bf r}\pm d\hat{y}) \cos (\omega \tau) \hat{P}_y (W_A({\bf r})e^{i\omega \tau})\sim \pm i \Omega$. Fourier transform of the tight binding model to the momentum space leads to the off-diagonal terms in equation (\ref{K}) with a momentum-dependent phase factor $\varphi_{\bf k}$. Near the $\Gamma$ point,  $\varphi_{\bf k}\sim arg\{-k_y+ik_x$\}. This corresponds to a two-dimensional pseudospin-orbit coupling $\sim k_y m_x+ k_xm_y$, where the sublattice index represents the pseudospin ${\bf m}$. It is known that, if one uses the Raman scheme, phase-locked multiple lasers are required for realizing a multi-dimensional spin-orbit coupling. Here, circularly driving the optical lattice itself naturally provides such an important pseudospin-orbit coupling. \\

{\bf Chiral d-wave superfluid induced by s-wave interaction}  Whereas the Hamiltonian in equation (\ref{K}) may lead to quantum anomalous Hall effect  in the strong  hybridization limit $4(t'_A-t'_B)-4|t_A-t_B|<\delta<4(t'_A-t'_B)+4|t_A-t_B|$,  we focus on the trivial single-particle band structure with Chern number $C=0$ in this Article. Strikingly, interaction completely changes the topological properties of the system, and gives rise to a chiral $d$-wave superfluid with a Chern number $C=2$. To see this fact, we consider two species of particles interacting through $s$-wave interaction, which are labeled as $\uparrow$ and $\downarrow$. We first diagonalize the single-particle Hamiltonian using,
\begin{equation}
\begin{split}
&\hat{c}^\dagger_{\bf k \sigma}=e^{i\varphi_{\bf k}}\beta_{\bf k }  \hat{a}^\dagger_{\bf k \sigma}-\alpha_{\bf k }\hat{b}^\dagger_{\bf k \sigma}\\
&\hat{d}^\dagger_{\bf k \sigma}=\alpha_{\bf k}\hat{a}^\dagger_{\bf k \sigma}+\beta_{\bf k}e^{-i\varphi_{\bf k} }\hat{b}^\dagger_{\bf k \sigma}\label{DE},
\end{split}
\end{equation}
where $\sigma=\uparrow, \downarrow$, $(\alpha_{\bf k}, \beta_{\bf k}e^{-i\varphi_{\bf k}})^T$ is the eigenstate of the matrix in equation (\ref{K}). Both $\alpha_{\bf k}$ and $\beta_{\bf k}$ are real, $a^\dagger_{\bf k}$ and $b^\dagger_{\bf k}$ are the creation operators for the unhybridized bands. Due to the phase winding $e^{i\varphi_{\bf k} }$ induced by the driving field, we refer the eigenstates as to helicity bands. Equation (\ref{DE}) avoids the ambiguity of the definition of the phase factor $\varphi_{\bf k}$ at the $\Gamma$ and $M$ points, where the off-diagonal term in equation (\ref{K}) vanishes and $(\alpha_{\bf k}, \beta_{\bf k})=(1,0)$. In other words, there is no band hybridization at these two points in the Brillouin zone (BZ). If one treats the sublattice index as a pseudospin, such a pseudospin points along the same direction at both the $\Gamma$ and M point in the same helicity band. Therefore, both bands are topologically trivial in the single-particle level with Chern number $C=0$. However, in the presence of interaction, a topologically nontrivial superfluid rises as the ground state.

\begin{figure}[tbp]
\begin{center}
\includegraphics[width=3.5in]{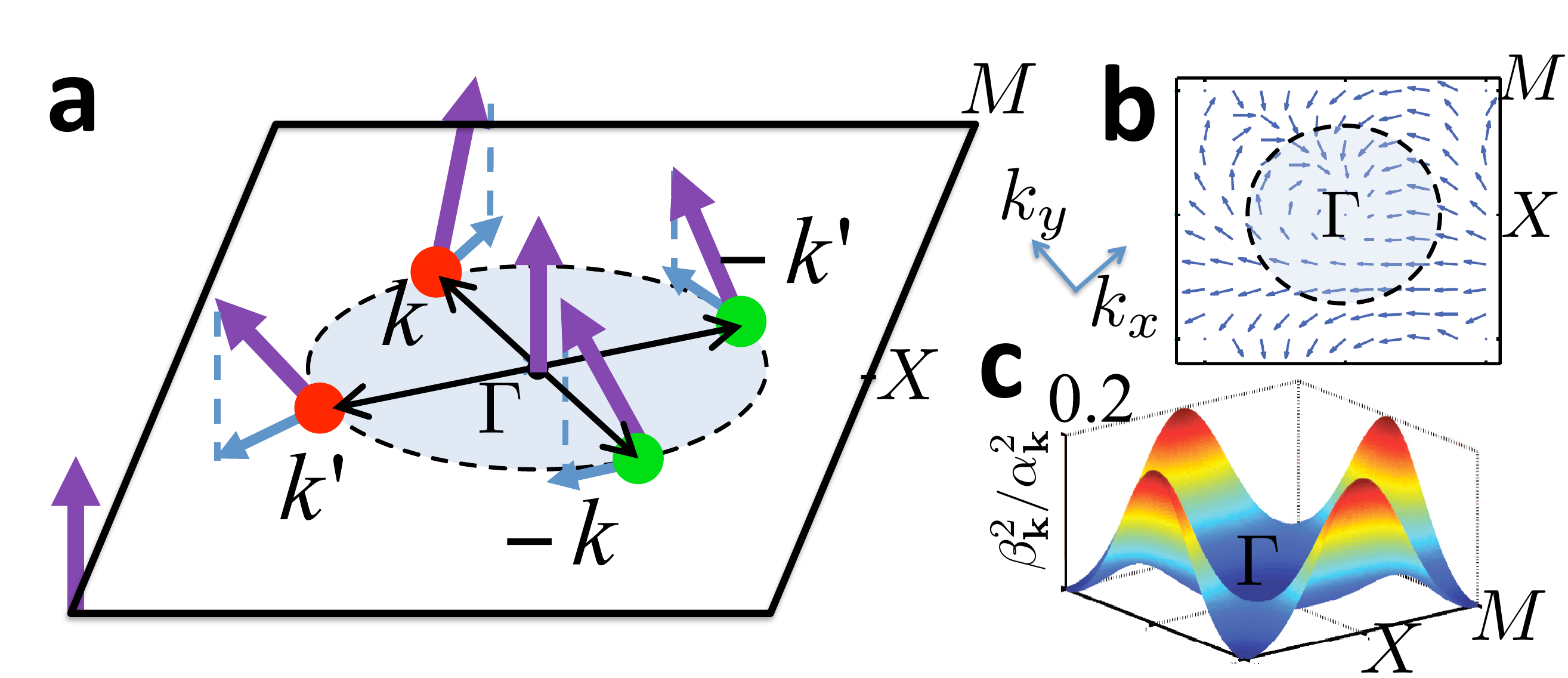}
\end{center}
\caption{\textbf{$d$-wave interaction induced by the phase winding in the single particle band structure.} (\textbf{a}) The purple arrows represent the pseudospin ${\bf m}$ formed by the $A$ and $B$ sublattices, and the blue arrows represent the projection of this pseudospin on the plane. At both the $\Gamma$ and $M$ points in the Brillouin zone, the pseudospin is perpendicular to the plane due to the absence of the band hybridization. The red and green spheres represent fermions with hyperfine spin up and spin down respectively. The dashed circle represents the Fermi surface and the shadowed region represents occupied states. Due to the phase winding in the single-particle spectrum, the scattering acquires a $d$-wave component that depends on the angle between the momentum ${\bf k}$ and ${\bf k}'$.  (\textbf{b}) Top view of (\textbf{a}). The projection of the pseudospin rotates $2\pi$ on the Fermi surface enclosing the $\Gamma$ point. (\textbf{c}) $\beta_{\bf k}^2/\alpha_{\bf k}^2$ in the Brillouin zone. $\beta_{\bf k}=0$ at the $\Gamma$ and M points. }
 \end{figure}

Consider two hyperfine spin states $\sigma=\uparrow, \downarrow$ uploaded to the shaken lattice, the interaction Hamiltonian containing only on-site attraction can be written as
\begin{equation}
\hat{V}=-U_A\sum_{i\in A}\hat{n}_{i\uparrow}\hat{n}_{i\downarrow}-U_B\sum_{i\in B}\hat{n}_{i\uparrow}\hat{n}_{i\downarrow},\label{int}
\end{equation}
where $\hat{n}_{i\sigma}$ is the density operator for spin-$\sigma$ particles at site $i$, and $U_{A,B}>0$. Without loss of generality, we have chosen $U_A\neq U_B$, since the lattice potentials near the minima of the A and B sublattice sites are in general different. For ultracold atoms, the ratio $U_A/U_B$ can be tuned in a broad range. First, for a fixed scattering length $a_s$($a_s<0$), $U_{\eta}=-\frac{2\pi\hbar^2a_s}{M}\int dz |\varphi(z)|^4\int d{\bf r}|\phi_{\eta}({\bf r})|^4$, where $\eta=A, B$, $\varphi(z)$ and $\phi_\eta({\bf r})$ are the Wannier wave functions of the lattice potential $V_z(z)$ and $V({\bf r})$ respectively.  As shown in figure 1, A sublattice sites have lower energies than B sublattice ones for a finite $s$, and therefore tighter confinements. The Wannier wave function $\phi_{A}({\bf r})$ is thus more localized than $\phi_{B}({\bf r})$. By tuning $s$, the ratio $U_A/U_B$ changes.  Second, this ratio can be further tuned by introducing a position dependent $a_s({\bf r})$ using optical Feshbach resonance, since a standing wave naturally leads to a periodically modulating scattering length $a_s(\bf r)$ in the real space\cite{OptFR}.  Alternatively, the current experimental advancements on creating magnetic lattices allow one to produce a spatially periodic magnetic field with largely tunable lattice geometry, period and amplitude\cite{ML1, ML2}. For a spatially dependent $a_s(\bf r)$, the on-site interaction then becomes $U_{\eta}=-\frac{2\pi\hbar^2}{M}\int dz |\varphi(z)|^4\int d{\bf r}a_s({\bf r})|\phi_{\eta}({\bf r})|^4$. It is even possible to strongly suppress one of them if $a_s({\bf r})$ is tuned to zero at the lattice sites of one of the sublattices.  Finally, there have been reports in a number of recent conferences that a well designed optical microscope generated by projecting a holographic mask \cite{Greiner} could create optical lattices with arbitrary lattice potentials, such as those going beyond traditional sinusoidal ones. Using this technique, $\phi_A({\bf r})$ and $\phi_B({\bf r})$ can be independent controlled and provide an extra degree of freedom to tune $U_A/U_B$. In this Article, we theoretically study how the ground state depends on $U_A/U_B$, and the driving field ${\bf A}(t)$. It is worth mentioning that all discussions here can be directly generalized to a related system, where the $A$ and $B$ sublattices are physically separated along the $z$ direction so that $\varphi(z)$ is different in A and B sublattices (Supplementary Note 1).

We first consider a filling factor $\langle \hat{n}_{i\uparrow} +\hat{n}_{i\downarrow} \rangle <2$. In the weakly interacting regime, where the interaction strength is much smaller than the band gap between the two helicity bands, we project $\hat{V}$ to the lower helicity band for constructing the low-energy effective theory.  We concrete the discussions on positive detuning $\delta>4(t'_A-t'_B)+4|t_A-t_B|$, since all results can be straightforwardly generalized to negative detuning $\delta<4(t'_A-t'_B)-4|t_A-t_B|$. For a positive detuning, after absorbing a photon, A sublattice sites still have lower energies than B sublattice ones. The main contribution to the lower helicity band comes from the A sublattice and the corresponding operator for the lower helicity band is $\hat{d}^\dagger_{\bf k}$. The BCS type of projected interaction Hamiltonian is written as $\hat{V}_{eff}=-\frac{1}{N^2}\sum_{\bf k k'}V_{\bf k k'}d^\dagger_{\bf k\uparrow}d^\dagger_{\bf -k\downarrow} d_{\bf -k' \downarrow}d_{\bf k'\uparrow}$, where
\begin{equation}
{V}_{\bf k k'}= \left(U_A \alpha_{\bf k}^2 \alpha_{\bf k'}^2 + U_B \beta_{\bf k}^2 \beta_{\bf k'}^2e^{i(\varphi_{\bf k}+\varphi_{\bf -k}-\varphi_{\bf k'}-\varphi_{\bf -k'})} \right)\label{Veff},
\end{equation}
where $N$ is the number of lattice sites. Equation (\ref{Veff}) is a key result of our work. It reveals a significant effect of the band hybridization in the driven lattice. Though the bare interaction is purely a $s$-wave one, the phase winding encoded in the two-dimensional pseudospin-orbit coupling $\sim k_y m_x+ k_xm_y$ leads to an effective interaction that contains both the $s$- and $d$-wave components. This can be intuitively understood from the fact that both the $\sigma=\uparrow$ and $\sigma=\downarrow$ components see the same driving field. Since the single-particle spectrum of each hyperfine spin state exhibits a phase winding $\sim e^{i\varphi_{\bf k}}$, the pairing scattering between $\sigma=\uparrow$ and $\sigma=\downarrow$ inevitably carries a $d$-wave component as shown in equation (\ref{Veff}).

The relative strengths of interaction in $s$- and $d$-partial-wave channels can be well controlled by microscopic parameters of the system, such as the ratio between the on-site interaction in A and B sublattices $U_A/U_B$. For the extreme cases,  $U_A=0$ or $U_B=0$, the interaction is a purely $s$- or $d$-wave one respectively.  It is expected that  the chiral $d$-wave superfluid shall emerge when $U_B$ is dominant.  Another important parameter is the ratio $\beta_{\bf k}/\alpha_{\bf k}$, which is controlled by the single-particle spectrum. At the $\Gamma$ and M points where the band hybridization is absent, $\beta_{\bf k}=0$ and $\alpha_{\bf k}=1$. The interaction is a purely $s$-wave one. Away from  the $\Gamma$ and M points,  $\beta_{\bf k}$ becomes finite. The larger the driving amplitude $A$ is, the larger the ratio $\beta_{\bf k}/\alpha_{\bf k}$ is. As shown later,  the value of $\beta_{\bf k}/\alpha_{\bf k}$ at the fermi energy is crucial to determine the topological properties of the superfluid.

\begin{figure}[tbp]
\begin{center}
\includegraphics[width=3.5in]{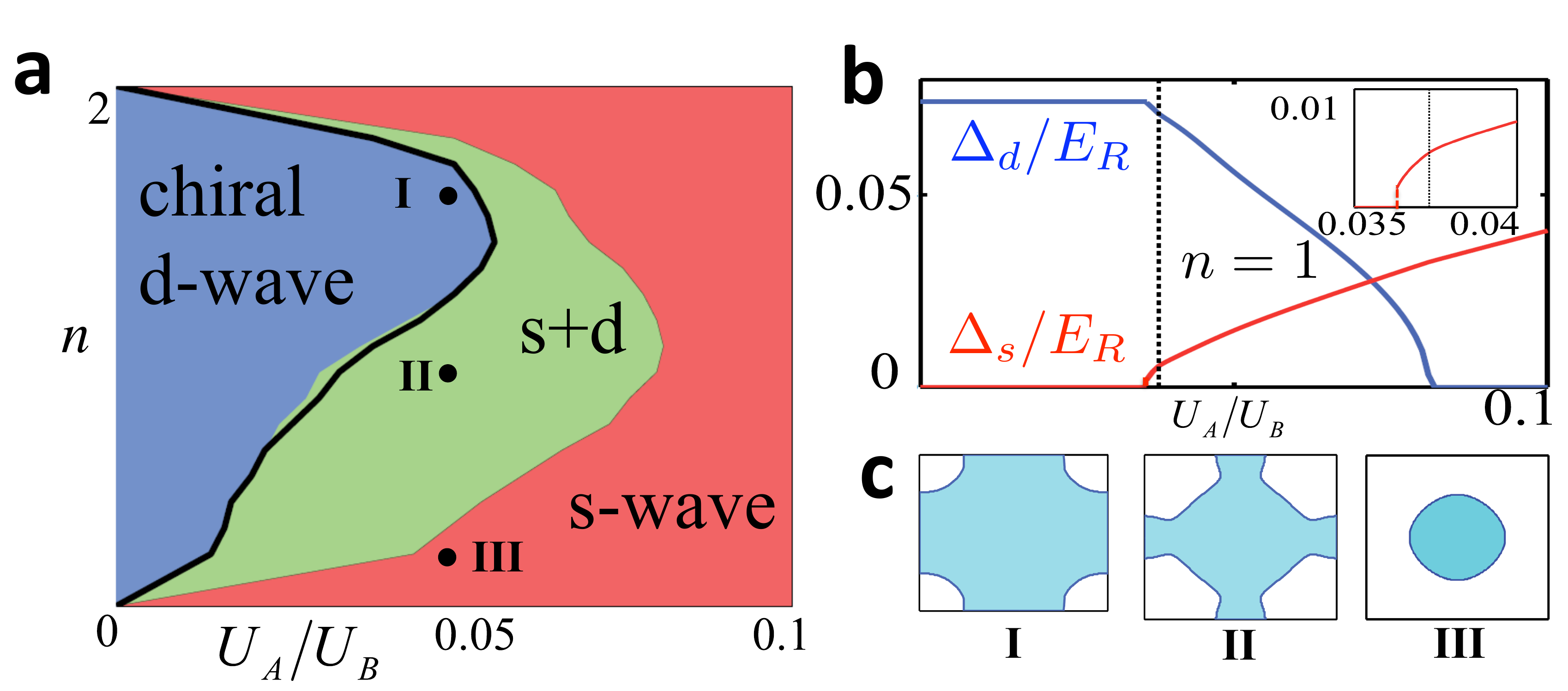}
\end{center}
\caption{\textbf{Phase diagram}. (\textbf{a}) The $x$ and $y$ axis represent the the ratio of the on-site interaction $U_A/U_B$ and the filling factor $n$ respectively. Chiral $d$-wave, $s$-wave and the mixed partial-wave superfluids are represented by blue, red and green colors respectively. For a fixed $n$, the chiral $d$-wave superfluid shows up at large enough $U_B$. The black curve represents the topological transition line characterizing the change of the Chern number from 2 to 0. For the numerical parameters we chose, it is very close to the boundary between the chiral $d$-wave and the $s+d$ superfluid. (\textbf{b}) $\Delta_s$ and $\Delta_d$ in terms of the recoil energy $E_R=\hbar^2\pi^2/(2Md^2)$ as a function of $U_A/U_B$ at half filling $n=1$. Inset shows the region near the topological phase transition point highlighted by the dashed vertical line.  (\textbf{c}) At a constant $U_A/U_B$, different filling factors lead to different superfluids. On different fermi surfaces, the contribution from the A and B sublattice to the helicity band, which is characterized by $\alpha_{\bf k}/\beta_{\bf k}$, is different, and different superfluids rise. (\textbf{I}), (\textbf{II}) and (\textbf{III}) represent the fermi surfaces of the $s$-wave, the mixed, and the chiral $d$-wave superfluids respectively before turning on the interaction.}
 \end{figure}


Define the order parameters in the $s$- and $d$-partial-wave channels,
\begin{equation}
\begin{split}
&\Delta_s=-U_A\sum_{\bf k}\alpha^2_{\bf k}\langle d_{-{\bf k}\downarrow}d_{{\bf k}\uparrow}\rangle/N,     \,\,\, \\
&\Delta_d=-U_B\sum_{\bf k}\beta^2_{\bf k}e^{-i(\varphi_{\bf k}+\varphi_{\bf -k})}\langle d_{-{\bf k}\downarrow}d_{{\bf k}\uparrow}\rangle/N ,
\end{split}
\end{equation}
the BCS mean field Hamiltonian can be written as
\begin{equation}
H_{M}=\frac{1}{N}\sum_{\bf k}\left(
\begin{array}{cc}
\epsilon_{\bf k}-\mu &\Delta_{s\bf k}+\Delta_{d{\bf k}} \\
\Delta^*_{s\bf k}+\Delta^*_{d{\bf k}}& \mu-\epsilon_{\bf k}
\end{array}\right),\label{BCS}
\end{equation}
where $\epsilon_{\bf k}$ and $\mu$ are the single-particle energy of the lower helicity band and the chemical potential respectively, $\Delta_{s\bf k}=\Delta_s\alpha_{\bf k}^2$, and $\Delta_{d{\bf k}}=\Delta_d \beta_{\bf k}^2e^{i(\varphi_{\bf k}+\varphi_{\bf -k})}$. At small momenta, $\beta_{\bf k}^2e^{i(\varphi_{\bf k}+\varphi_{\bf -k})}\sim (k_x+ik_y)^2$.
Thus $\Delta_d$ corresponds to the order parameter in the $d_{x^2-y^2}+id_{xy}$ channel. $\Delta_s$ and $\Delta_d$ can be obtained self-consistently through the standard BCS approach (Supplementary Note 2), and the ground state wave function is,
\begin{equation}
|G\rangle=\prod_{\bf k\in BZ}(u_{\bf k}+v_{\bf k} e^{i\theta_{\bf k}}\hat{d}^\dagger_{\bf k \uparrow }\hat{d}^\dagger_{\bf -k \downarrow})|0\rangle,\label{gs}
\end{equation}
where $u_{\bf k}$ and $v_{\bf k}$ are real, and $\theta_{\bf k}=\arg\{\Delta_s\alpha_{\bf k}^2+\Delta_d \beta_{\bf k}^2e^{i(\varphi_{\bf k}+\varphi_{\bf -k})}\}$. Microscopically, $\Delta_s$ and $\Delta_d$ are controlled by  $U_A/U_B$, and the parameters of single-particle Hamiltonian, including both the detuning $\delta$ and the strength of the driving field ${\bf A}$. Two extreme cases are rather clear. When  $U_B=0$, $\Delta_d$ is always zero, and $\theta_{\bf k}=0$. The ground state is a conventional $s$-wave superfluid, $|G\rangle_s=\prod_{\bf k}(u_{\bf k}+v_{\bf k} \hat{d}^\dagger_{\bf k \uparrow }\hat{d}^\dagger_{\bf -k \downarrow})|0\rangle$. In contrast, when $U_A=0$,  $\Delta_s$ vanishes, and the superfluid is $d_{x^2-y^2}+id_{xy}$ superfluid as expected, since the interaction in equation (\ref{Veff}) is a purely $d$-wave one. The ground state becomes $|G\rangle_d=\prod_{\bf k}(u_{\bf k}+v_{\bf k} e^{i(\varphi_{\bf k}+\varphi_{\bf -k})}\hat{d}^\dagger_{\bf k \uparrow }\hat{d}^\dagger_{\bf -k \downarrow})|0\rangle$. Using the pseudospin representation where $d^\dagger_{\bf k \uparrow }d^\dagger_{\bf -k \downarrow}|0\rangle$  and $|0\rangle$ are treated as the up and down component of a pseudospin $S_z=\pm 1/2$,  Bloch sphere is covered twice by this pseudospin defined in the Brillouin zone, i.e., the Chern number is 2, a characteristic feature of topological superfluids.


For general values of $U_A$ and $U_B$, the phase diagram is shown in figure 3. For any given single-particle spectrum with a fixed $\delta$ and ${\bf A}$,  the chiral $d$-wave superfluid is the ground state for large enough $U_B/U_A$. For a spatially dependent scattering length, we use the ansatz $a_s({\bf r})=a_0-a_1 \cos (\pi x/d)\cos(\pi y/d)$ to model it. For $a_0\sim 20 nm$, and $a_1\sim 16 nm$, which corresponds to $U_A/U_B$ about $0.05-0.1$, we find out that $\Delta_d$ can be a few to ten $nK$. With further increasing $U_B$, $\Delta_d$ could be even larger.  With the current advancements of cooling atoms \cite{Ketterlecooling, Hulet, Coolingreview}, it is promising to access such a temperature scale in experiments. For a uniform scattering length, one could also adjust $U_B/U_A$ and access the chiral $d$-wave superfluid by adjusting the lattice potential itself via changing $V$ and $s$ (Supplementary Note 3). It is worth mentioning that, in an ordinary checkerboard or square lattice that respects the $D_4$ symmetry, it is difficult to couple $d_{x^2-y^2}$ and $d_{xy}$ which belong to different irreducible representations of the $D_4$ group. In our system, the time-reversal symmetry is readily broken in the single-particle level so that such a constraint is absent. In the weakly interacting regime, particles prefer to stay in the lower helicity band, where the effective interaction inevitably includes a $d_{x^2-y^2}+id_{xy}$ component, and the chiral $d$-wave superfluid naturally rises.

 With increasing $U_A/U_B$, there is a first order transition from a pure chiral $d$-wave superfluid  to a mixed $s$- and $d$-wave superfluid in the mean field framework.  Such a mixed superfluid may be referred as to $s+(d_{x^2-y^2}+id_{xy})$. Whereas both a pure $s$ and a pure chiral $d$-wave superfluids are gapped, such a mixed state may be gapless by fine tuning of the parameters. When the gap closes, it corresponds to a topological phase transition point defined by
\begin{equation}
\Delta_s\alpha_{\bf k^*}^2=\Delta_d \beta_{\bf k^*}^2,\,\,\,\,\, \epsilon_{\bf k^*}=\mu \label{cv}
\end{equation}
where ${\bf k}^*$ is a momentum on the Fermi surface along the diagonal direction of BZ such that $k^*_x=0$.  Along this direction, the off-diagonal term in the BCS Hamiltonian becomes real and could vanish, since $\varphi_{\bf k^*}+\varphi_{-\bf k^*}=\pi$. When equation (\ref{cv}) is satisfied, both the diagonal and off-diagonal term of the BCS Hamiltonian in equation (\ref{BCS}) vanish so that the excitation gap disappears.  Across the transition point, and the Chern number of the ground state jumps from 2 to 0. On each side of the transition point, the mixed state is topologically equivalent to either the chiral $d_{x^2-y^2}+id_{xy}$ or the $s$-wave superfluid (Supplementary Note 4). It is worth mentioning that for the mixed $s$- and $d$-wave superfluid, there is a two-fold degeneracy, since changing the sign of $\Delta_d$ in equation (\ref{BCS}) leads to the same ground state energy. For such a $s-(d_{x^2-y^2}+id_{xy})$ superfluid, ${\bf k}^*$ satisfies $k^*_y=0$ such that $\varphi_{\bf k^*}+\varphi_{-\bf k^*}=0$.


Equation (\ref{cv}) tells one that, in addition to changing $U_A/U_B$, tuning the filling factor or the driving field also allows one to access the topological phase transition, since the amplitude of the driving field and the location of the Fermi surface  determine the value of $\beta_{\bf k_F}/\alpha_{\bf k_F}$. As shown in figure 3,  for a Fermi surface very close to the $\Gamma$ or the M point, $\beta_{\bf k_F}/\alpha_{\bf k_F}$ is small, and the $s$-wave interaction is dominate so that the Chern number is 0. With a constant $U_B/U_A$, enlarging the Fermi surface or increasing the driving amplitude ${\bf A}$ increases $\beta_{\bf k_F}/\alpha_{\bf k_F}$ and leads to the transition to the ground state with $C=2$.

A characteristic feature of the chiral $d$-wave superfluid is the presence of  two chiral edge states, as shown in figure 4, which serve as unique signatures of the chiral $d$-wave superfluids in experiments. We have chosen the edge along the $x'$ direction cutting A sublattices so that $k_x'\sim k_x+k_y$ is a good quantum number. Along the $y'$ direction, we have applied the open boundary condition. In solid materials, such chiral edge states correspond to a quantized Hall conductivity $\sigma_{H}=2e^2/h$. In ultracold atoms, such edge states could also be measured using a variety of techniques\cite{edge0}, such as the expansion of the atomic cloud\cite{edge1}, and the generalization of the Bragg spectroscopy that is sensitive to angular momentum \cite{edge2}. This allows one to directly visualize the edge states and identify the chiral $d$-wave superfluid.  Such  edge states are topologically protected, since a small mixing with the $s$-wave component in the $s+(d_{x^2-y^2}+id_{xy})$ superfluid retains the Chern number $C=2$.

\begin{figure}[tbp]
\begin{center}
\includegraphics[width=3.6in]{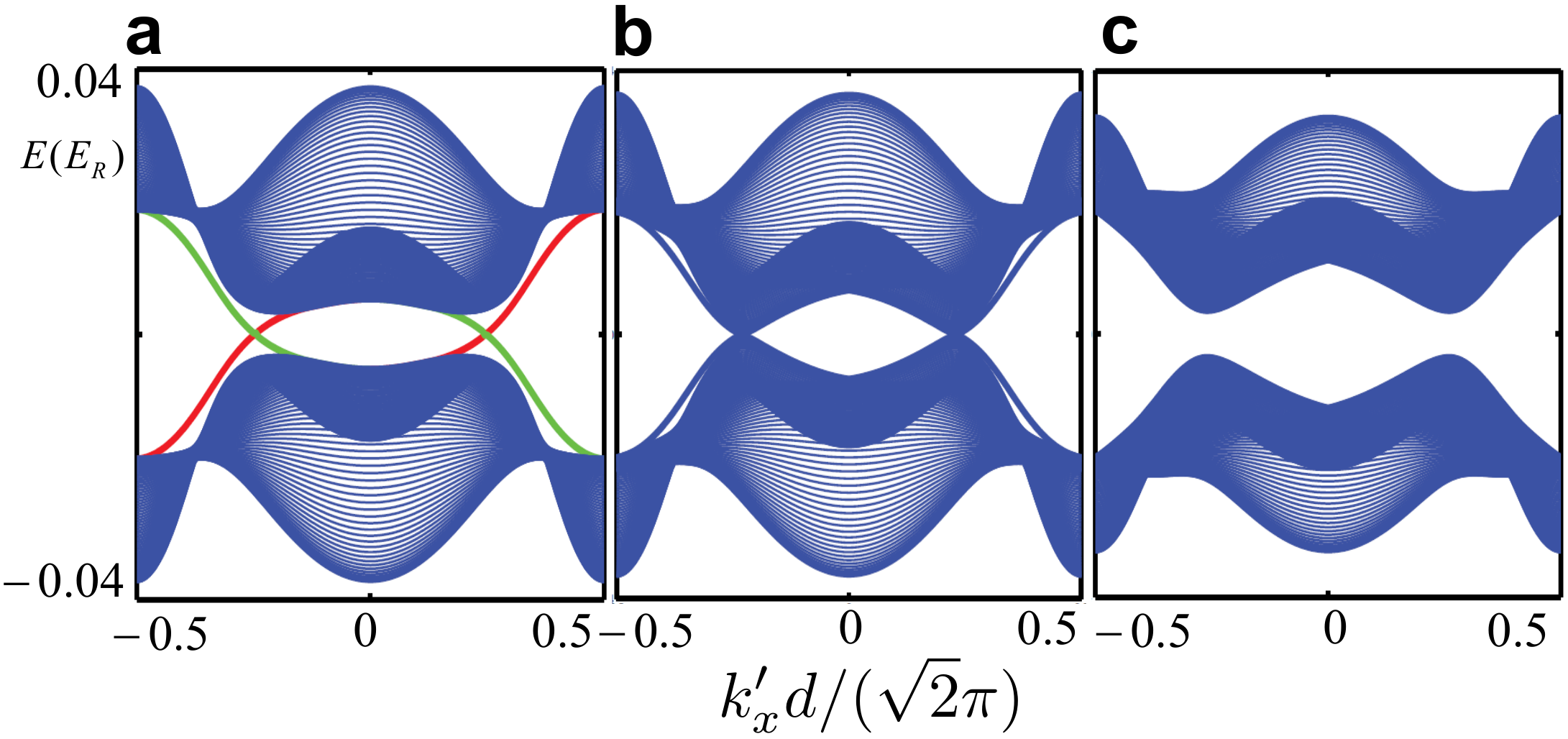}
\end{center}
\caption{\textbf{Edge states}.  (\textbf{a}) When $U_A/U_B$ is small, the  topologically non-trivial ground state has two chiral edge states on each edge(red and green solid lines). (\textbf{b}) When $U_A/U_B$ increases to a critical value, the gap in the excitation spectrum vanishes. A topological phase transition occurs, and edge states disappear. (\textbf{c}) With further increasing  $U_A/U_B$, the gap reopens in the bulk and  there is no edge state in the topologically trivial state. }
 \end{figure}

For a large filling factor $\langle \hat{n}_{i\uparrow} +\hat{n}_{i\downarrow} \rangle>2$, the fermi surface shall form in the upper helicity band. One then needs to project the interaction to the upper helicity band, and $\hat{V}_{eff}'=-\frac{1}{N^2}\sum_{\bf k k'}V'_{\bf k k'}c^\dagger_{\bf k\uparrow}c^\dagger_{\bf -k\downarrow} c_{\bf -k' \downarrow}c_{\bf k'\uparrow}$, where
\begin{equation}
{V}_{\bf k k'}'= \left(U_B \alpha_{\bf k}^2 \alpha_{\bf k'}^2  +U_A \beta_{\bf k}^2 \beta_{\bf k'}^2  e^{i(\varphi_{\bf k'}+\varphi_{\bf -k'}-\varphi_{\bf k}-\varphi_{\bf -k})} \right)\label{Veffp},
\end{equation}
similar to $V_{eff}$ for the lower helicity band,  with inverted roles of $U_B$ and $U_A$, as well as those of $\alpha_{\bf k}$  and $ \beta_{\bf k} $.  All previous discussions then apply. \\

{\bf Discussions} To reveal the underlying physics for creating the chiral $d$-wave superfluid, we have projected the interaction to one of the helicity band. Whereas this is well justified in the weak interacting limit, it is useful to investigate the interaction induced mixing between the two helicity bands.  We write the full Hamiltonian as $H=\sum_{\bf k}\hat{\Psi}^\dagger_{\bf k} M_o \Psi_{\bf k}$, where $\hat{\Psi}^\dagger_{\bf k}=(a^\dagger_{\bf k \uparrow}, b^\dagger_{\bf k \uparrow }, a_{-\bf k \downarrow}, b_{-\bf k \downarrow})$
\begin{equation}
M_o=\left(\begin{array}{cccc} \epsilon_{A{\bf k}}-\mu & \Omega_{\bf k}e^{i\varphi_{\bf k}} & \Delta_A & 0 \\ \Omega_{\bf k}e^{-i\varphi_{\bf k}} & \epsilon_{B{\bf k}}-\mu & 0 & \Delta_B \\ \Delta^*_A & 0 & \mu-\epsilon_{A{\bf k}} & -\Omega_{\bf k}e^{-i\varphi_{\bf -k}} \\ 0 & \Delta^*_B & -\Omega_{\bf k}e^{i\varphi_{\bf -k}} & \mu-\epsilon_{B{\bf k}} \end{array}\right), \label{4by4}
\end{equation}
where $\epsilon_{\eta{\bf k}}$ is the single-particle energy of $\eta=A, B$ sublattices,  $\Delta_A=-U_A\sum_{\bf k}\langle a_{\bf k \uparrow} a_{\bf -k \downarrow}\rangle/N $ and $\Delta_B=-U_B\sum_{\bf k}\langle b_{\bf k \uparrow} b_{\bf -k \downarrow}\rangle/N$ characterize the on-site pairing amplitudes. For noninteracting systems, equation (\ref{4by4}) describes two identical copies of equation (\ref{K}) for spin-up and spin-down particles. The presence of attraction leads to the particle-hole hybridization in both A and B sublattices. Using the basis of  single-particle eigenstates as defined in equation (\ref{DE}),  the Hamiltonian can be rewritten as $H=\sum_{\bf k}\hat{\Phi}^\dagger_{\bf k} M_h \Phi_{\bf k}$, where $\hat{\Phi}^\dagger_{\bf k}=(d^\dagger_{\bf k \uparrow}, d_{-\bf k \downarrow }, c^\dagger_{\bf k \uparrow}, c_{-\bf k \downarrow})$  and
\begin{equation}
M_h=\left(\begin{array}{cccc}
\epsilon_{d{\bf k}}-\mu & \Omega_{d{\bf k}} & 0 & \nu_{\bf k} \\ \Omega^\ast_{d{\bf k}} & \mu-\epsilon_{d{\bf k}} & \nu^\ast_{\bf -k} & 0 \\  0 & \nu_{\bf -k} & \epsilon_{c{\bf k}}-\mu & \Omega_{c{\bf k}} \\ \nu^\ast_{\bf k} & 0  & \Omega^\ast_{c{\bf k}} & \mu-\epsilon_{c{\bf k}}
\end{array}\right), \label{dre}
\end{equation}
which includes both helicity bands where $\Omega_{d{\bf k}}=\Delta_A\alpha^2_{\bf k}+\Delta_B\beta^2_{\bf k}e^{i(\varphi_{\bf k}+\varphi_{\bf -k})}$, $\Omega_{c{\bf k}}=\Delta_B\alpha^2_{\bf k}+\Delta_A\beta^2_{\bf k}e^{-i(\varphi_{\bf k}+\varphi_{\bf -k})}$ and $\nu_{\bf k}=\alpha_{\bf k}\beta_{\bf k}(\Delta_Ae^{-i\varphi_{\bf -k}}-\Delta_Be^{i\varphi_{\bf k}})$. For weak interactions that are much smaller than the gap between the two helicity bands, the block off-diagonal matrix can be ignored, and the upper and lower helicity bands can be investigated independently, as discussed before. It is easy to identify that
$\Delta^c_s= \Delta_B$, $\Delta^c_d= \Delta_A$, $\Delta^d_s= \Delta_A$, $\Delta^d_d= \Delta_B$ where the superscript $c$ and $d$ represent the results for two different helicity bands respectively.

To include the effect of inter-helicity band coupling, we have numerically solved equation (\ref{dre}), and verified that the mixing of the upper helicity band is exponentially small in the weak interacting limit. We have also verified that both the Chern number of ground state and the existence of the chiral edge states are not affected neither. This proves that the chiral $d$-wave superfluid discovered in this Article is stable against a weak inter-helicity band coupling.

The study on topological matters is an important topic of current studies in both condensed matter and cold atom physics. Whereas a number of non-interacting topological matters, such as topological insulators, have been realized in experiments \cite{HgTe,BiSb,BiSb2}, the search for topological superfluids, which naturally encode the interaction effects, has just started \cite{Schnyder,Kitaev,Qi,Qi2}. This Article has proposed a new principle for realizing a chiral $d$-wave superfluid without resorting to a strong $d$-wave interaction. With the induced $d$-wave interaction in the driven lattice, a bare $s$-wave interaction is sufficient to create a chiral $d$-wave superfluid. Since periodic driving fields allow one to engineer the Bloch-Floquet bands, it is expected that applying such driving fields provides physicists unique opportunities to create a variety of synthetic high-partial-wave interactions. It is promising that even more exotic topological superfluids, as well as other interacting topological matters, shall be created.

Our scheme also provides physicists an ideal platform for studying the interplay between the topology of single-particle band structures and interaction. Whereas significant progresses have been made in understanding topological matters in non-interacting systems, many questions remain open when interaction is introduced. This article has focused on a case where the single-particle band structure is topologically trivial, with Chern number $C=0$. We have seen that the interplay between interaction and a special band structure that encodes a phase winding in the single-particle eigenstates gives rise to topologically non-trivial matters with Chern number $C=2$.  Since the topology of the band structure in our system can be well controlled, we believe that this work will stimulate more studies on the interaction effect in a topologically nontrivial band structure of a periodically driven lattice.

{\bf Acknowledgement }
QZ acknowledges useful discussions with T. Esslinger and K.T. Law. This work was supported by ECS/RGC(409513) and CRF/RGC(HKUST3/CRF/13G).

{\bf Author Contributions }
QZ conceived the project. SLZ and LJL performed  the  analytical and numerical studies. QZ wrote the paper.

{\bf  Competing financial interests}  The authors declare no competing financial interests.

{\bf  Corresponding author} Correspondence and requests for materials should be addressed to qizhou@phy.cuhk.edu.hk

\onecolumngrid


\newpage

\centerline{\bf Supplementary Figures}

\begin{figure}[h]
\begin{center}
\includegraphics[width=3in]{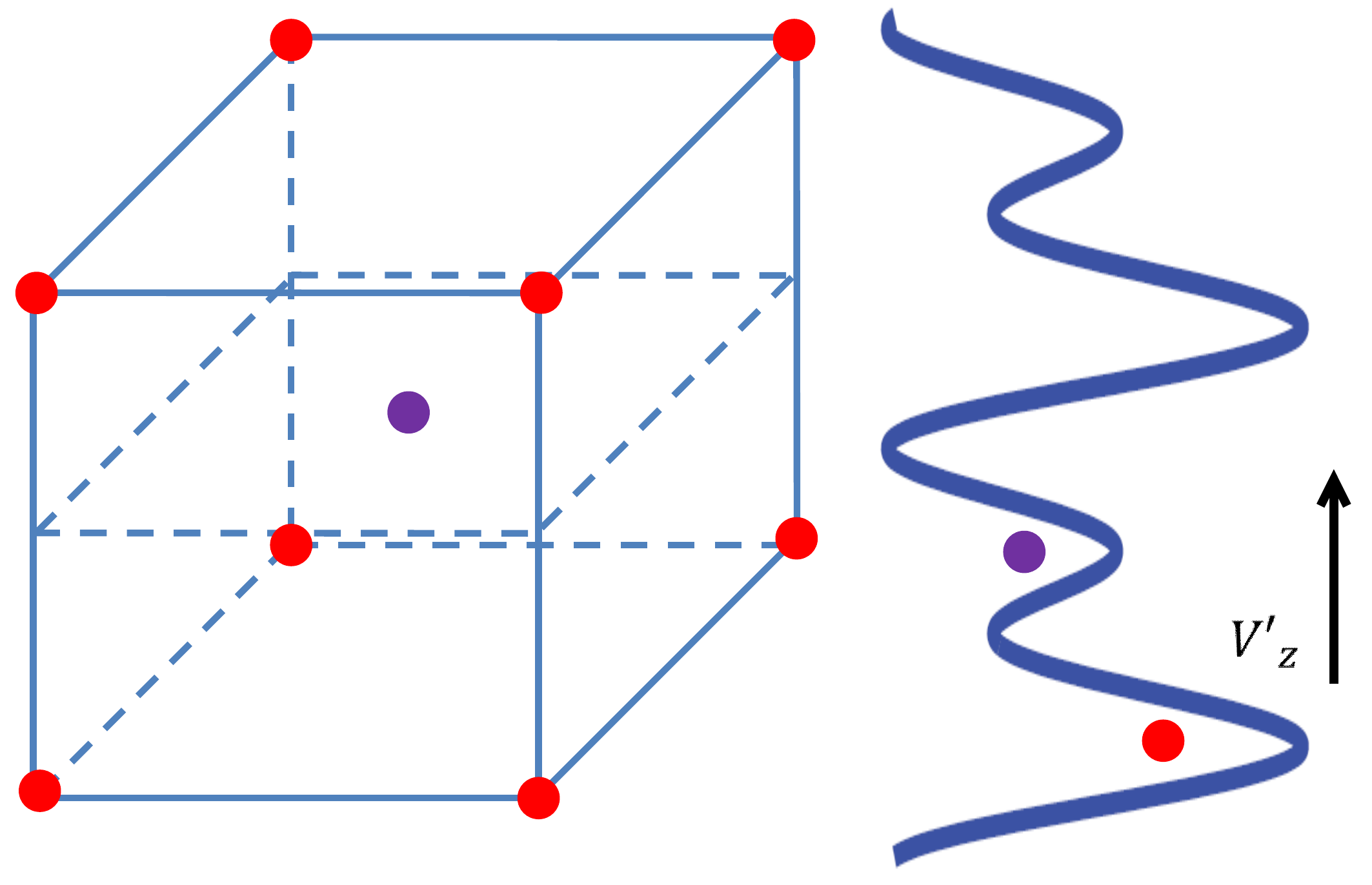}
\end{center}
\caption{ \textbf{A body-centered cubic  lattice with a tunable $U_A/U_B$ }. An additional lattice potential along the  $z$ direction differentiates the lattices sites at the corner (red dots ) and the center (purple dot) of each cube, which form the A and B sublattices.  In such a lattice, the lattice potential in different layers along the $z$-direction is different, and the ratio $U_A/U_B$ depends on the amplitude $V_z'$. For large enough $V_z'$, such a three-dimensional lattice is decomposed to many independent two-dimensional systems, each of which is a checkerboard lattice with A and B sublattice shifted along the $z$ direction.}
 \end{figure}

\begin{figure}[h]
\begin{center}
\includegraphics[width=6in]{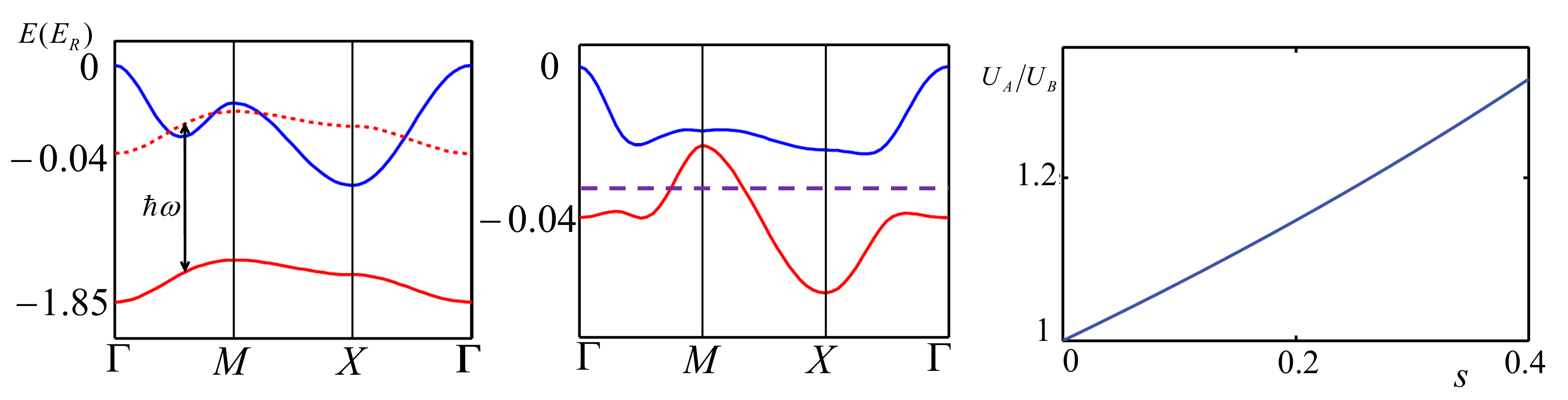}
\end{center}
\caption{ \textbf{Realize chiral $d$-wave superfluid with a uniform scattering length}. Left,  The lowest two bands (red and blue solid curve) of the static lattice with the parameter $V=6E_R$, $s=0.1$. Red dashed line is the side band formed by the $A$ band with one photon absorption. Middle,  Floquet band structure after the lattice is periodically driven. Purple dashed line is the fermi energy.  In this case, with a fixed scattering length $a_s=9.2nm$, the ground state is a topologically non-trivial $s+d$ wave superfluid. Right: the dependence of $U_A/U_B$ on $s$ with a fixed scattering length. }
 \end{figure}

\begin{figure}[h]
\begin{center}
\includegraphics[width=5in]{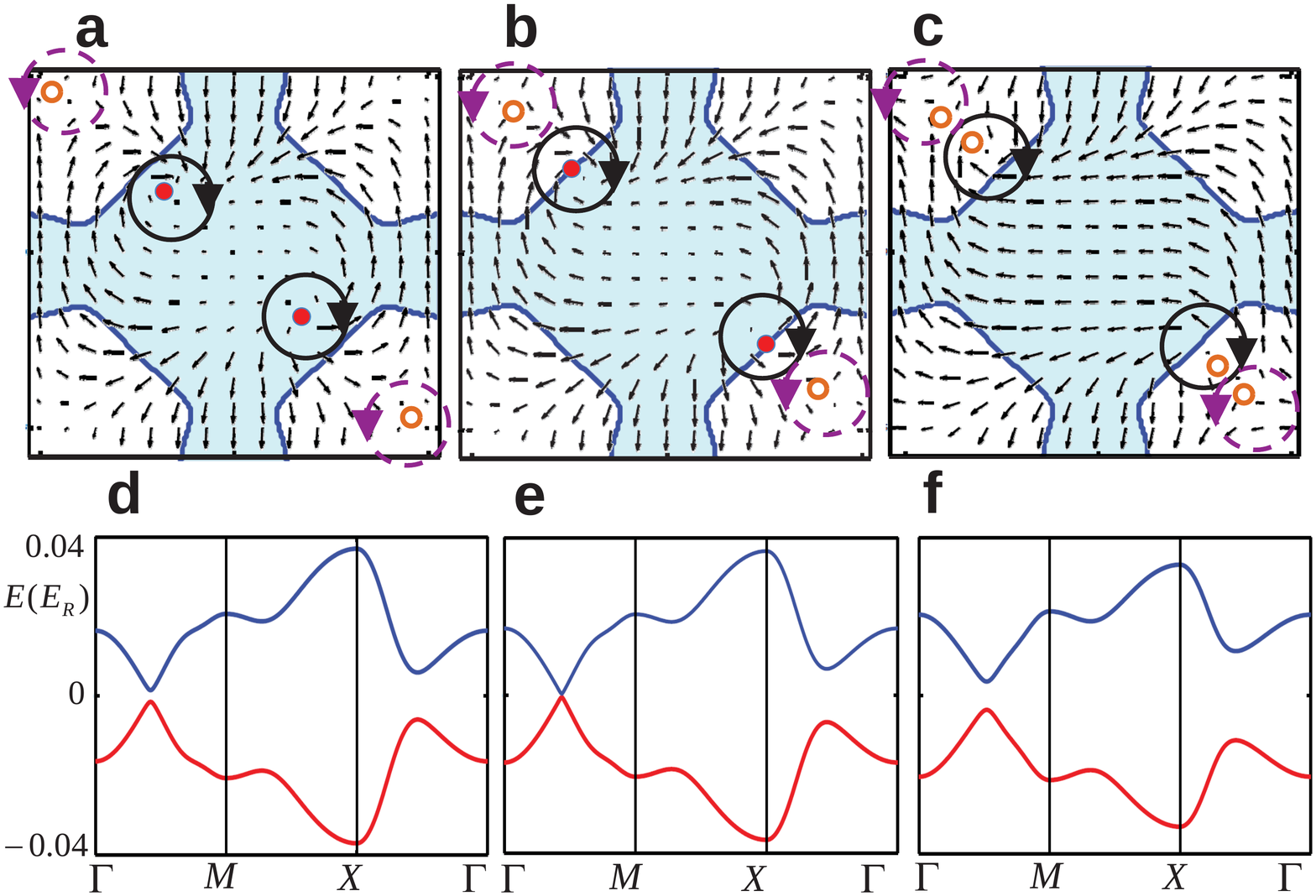}
\end{center}
\caption{ \textbf{Topological phase transition.} (\textbf{a,b,c}) Small arrows characterize the strength and direction of the $(S_x, S_y)$ component of the pseudospin, which represents the particle and hole states in the BCS wave function, in the $s+d$ superfluid. Small filled and empty dot represent the location in the BZ where the pseudospin has only the $S_z$ component with $S_z=1/2$ and $S_z=-1/2$ respectively. Along the black solid and purple dashed loop, the $(S_x, S_y)$ component of the pseudospin has winding number $1$ and $-1$. Blue solid line is the fermi surface and the shadowed region represents the occupied state. In (\textbf{a}), $\Delta_s$ is very small, two filled ($S_z=1/2$) and two empty dots($S_z=-1/2$) are present in BZ. The resultant superfluid is topological non-trivial($C=2$), since the Bloch sphere is covered twice by the pseudospin. (\textbf{b}) When $\Delta_s$ increases, the two filled dots move from $\Gamma$ to $M$ and touch the fermi surface, this is the critical point of topological phase transition, where the excitation gap vanishes at the positions of the red filled dots on the Fermi surface. (\textbf{c}) Further increasing $\Delta_s$,  no point in the BZ has a pseudospin state $S_z=1/2$, i.e., the north pole of the Bloch sphere is not covered, and the superfluid is topological trivial($C=0$). (\textbf{d,e,f}) Corresponding excitation spectrum of the $s+d$ wave superfluid. (\textbf{e}) The excitation becomes gapless on the fermi surface at the critical point of topological phase transition.}
 \end{figure}

\newpage

\centerline{\bf Supplementary Note 1. An alternative scheme to realize a bipartite lattice with tunable $U_B/U_A$}
\vspace{0.1in}

\par
We consider a three-dimensional lattice, which may be created by four pairs of lasers,
\begin{equation}
V({\bf r})=-V_0\left(\cos^2(\frac{\pi x}{d})+\cos^2(\frac{\pi y}{d})\right)-V({z})-sV_0 \cos(\frac{\pi x}{d})\cos(\frac{\pi y}{d})-s'\sqrt{V_0V_z}\left(\cos(\frac{\pi y}{d})\cos(\frac{\pi z}{d})+\cos(\frac{\pi x}{d})\cos(\frac{\pi z}{d})\right), \label{bccl}
\end{equation}
where $V(z)=V_z\cos^2(\frac{\pi z}{d})-V'_z\cos(\frac{\pi z}{d})$ is a one-dimensional double well lattice along the $z$ direction. When $V'_z=0$, such a lattice is a body centered cubic lattice. A large enough $V'_z$ separates the system into decoupled two-dimensional ones, each of which is composed of two square lattices separated along the $z$ direction by a distance $d$, as shown in Fig.S1. View from the top, such a two-dimensional lattice is equivalent to a checkerboard lattice.
Whereas all the results for both non-interacting and interacting systems are the same as the checkerboard lattice discussed in the main text, such a lattice has an advantage in practice. Since the A and B sublattices are readily separated along the $z$ direction, a scattering length $a_s(z)$ that changes only along the $z$ direction is sufficient, if one wants to use such a scheme to further change $U_A/U_B$.


\vspace{0.1in}


\centerline{\bf  Supplementary Note 2. Self-consistent solutions of the BCS Hamiltonian}
\vspace{0.1in}

\par
Define the order parameter:
\begin{equation}
\Delta_s=-U_A\sum_{\bf k}\alpha^2_{\bf k}\langle d_{-{\bf k}\downarrow}d_{{\bf k}\uparrow}\rangle/N,     \,\,\, \Delta_d=-U_B\sum_{\bf k}\beta^2_{\bf k}e^{-i(\varphi_{\bf k}+\varphi_{\bf -k})}\langle d_{-{\bf k}\downarrow}d_{{\bf k}\uparrow}\rangle/N
\end{equation}
The mean-field interaction Hamiltonian is:
\begin{eqnarray}
H_I&=&-\frac{1}{N^2}\sum_{{\bf k}{\bf k}'}V_{{\bf k}{\bf k}'}\left(\langle d^\dag_{{\bf k}\uparrow}d^\dag_{-{\bf k}\downarrow}\rangle d_{-{\bf k}'\downarrow}d_{{\bf k}'\uparrow}+d^\dag_{{\bf k}\uparrow}d^\dag_{-{\bf k}\downarrow}\langle d_{-{\bf k}'\downarrow}d_{{\bf k}'\uparrow}\rangle-\langle d^\dag_{{\bf k}\uparrow}d^\dag_{-{\bf k}\downarrow}\rangle\langle d_{-{\bf k}'\downarrow}d_{{\bf k}'\uparrow}\rangle\right)
                                    \nonumber\\
&=&\frac{1}{N}\left(\Delta^\ast_s\sum_{\bf k}\alpha^2_{\bf k}d_{-{\bf k}\downarrow}d_{{\bf k}\uparrow}+\Delta^\ast_d\sum_{\bf k}\beta^2_{\bf k}e^{-i(\varphi_{\bf k}+\varphi_{\bf -k})} d_{-{\bf k}\downarrow}d_{{\bf k}\uparrow}+h.c.\right)+\left(\frac{|\Delta_s|^2}{U_B}+\frac{|\Delta_d|^2}{U_A}\right)
\end{eqnarray}
Define $\Omega_{\bf k}e^{i\theta_{\bf k}}=\Delta_s\alpha^2_{\bf k}+\Delta_d\beta^2_{\bf k}e^{i(\varphi_{\bf k}+\varphi_{\bf -k})}$ and $\xi_{\bf k}=\epsilon_{\bf k}-\mu$ where $\mu$ is the chemical potential. Using Bogoliubov transformation, one can get the self-consistent equation about the order parameter:
\begin{equation}
\Delta_s=\frac{U_A}{2N}\sum_{\bf k}\alpha^2_{\bf k}\frac{\Delta_s\alpha^2_{\bf k}+\Delta_d\beta^2_{\bf k}e^{i(\varphi_{\bf k}+\varphi_{\bf -k})}}{\sqrt{\xi^2_{\bf k}+|\Delta_s\alpha^2_{\bf k}+\Delta_d\beta^2_{\bf k}e^{i(\varphi_{\bf k}+\varphi_{\bf -k})}|^2}}
\label{eq:orders}
\end{equation}
\begin{equation}
\Delta_d=\frac{U_B}{2N}\sum_{\bf k}\beta^2_{\bf k}\frac{\Delta_s\alpha^2_{\bf k}e^{-i(\varphi_{\bf k}+\varphi_{\bf -k})}+\Delta_d\beta^2_{\bf k}}{\sqrt{\xi^2_{\bf k}+|\Delta_s\alpha^2_{\bf k}+\Delta_d\beta^2_{\bf k}e^{i(\varphi_{\bf k}+\varphi_{\bf -k})}|^2}}
\label{eq:orderd}
\end{equation}
This set of equations is solved numerically for obtaining the results presented in the main text.

\vspace{0.1in}

\centerline{\bf  Supplementary Note 3. Tuning $U_A/U_B$ with a uniform scattering length}
\vspace{0.1in}

\par


In the main text, we have discussed how to tuning $U_A/U_B$ using a spatially variant scattering length $a_s({\bf r})$ for accessing the chiral d-wave superfluid. Even with a uniform scattering length $a_s$, the same goal can be achieved, since
the phase transition from $s$ to $d$-wave superconductor also depends on $\beta_{\bf k}/\alpha_{\bf k}$.
By engineering the single-particle band structure so that 
$\beta_{\bf k}$ is much larger than $\alpha_{\bf k}$  at the Fermi surface, the $d$-wave interaction becomes dominant so that the chiral $d$-wave superconductor emerges.  For the parameters shown in supplementary figure S2, we have found out that when the filling factor $\langle \hat{n}_{i\uparrow} +\hat{n}_{i\downarrow} \rangle=1.6$, $\Delta_d=0.01E_R$.
\par



\vspace{0.2in}

\centerline{\bf Supplementary Note 4. Chern number of the $s\pm (d_{x^2-y^2}+ id_{xy})$ superfluid}
\vspace{0.1in}

\par

For a pure $d$-wave superfluid with $\Delta_s=0$, the mapping from the BZ to the Bloch sphere gives rise to a Chern number of $2$. The $\Gamma$ and $M$ point correspond to the north and south pole, if one views the occupied and unoccupied states as spin-up and spin-down state of a pseudospin-1/2.  This can be seen directly from equation (9) of the main text, since $\beta_{\bf k}=0$ at the $\Gamma$ and $M$ point.

When $\Delta_s$ becomes finite, there will be four points in the BZ that lead to vanishing off-diagonal terms of the matrix in equation (9) of the main text for a given ratio $\Delta_s/\Delta_d$. For $s\pm (d_{x^2-y^2}+ id_{xy})$ superfluid, this corresponds to $\Delta_s\alpha_{\bf k}^2=\pm \Delta_d \beta_{\bf k}^2$. As shown in supplementary figure 3, if none of these four points touches the Fermi surface, the spectrum remains fully gapped, since the matrix is always finite. Moreover, it is easy to see that the Bloch sphere is still covered twice, and the Chern number remains to be 2. Under this situation, the mixed superfluid is topologically equivalent to a chiral $d$-wave superfluid. With increasing $\Delta_s$, two of these four points will touch the Fermi surface, and at these two points, the excitation spectrum becomes gapless. If $\Delta_s$ further increases, all these four points will be located either within or outside the Fermi surface, so that either the south or the north pole cannot be covered. This  means that the Chern number becomes zero, and the mixed superfluid is equivalent to a trivial $s$-wave superfluid.

\end{document}